# CNoA: Challenging Number Approach for uncovering TCP SYN flooding using SYN spoofing attack

L.Kavisankar[1] and C.Chellappan[2]

[1,2]Department of Computer Engineering, Anna University, Chennai, Tamil Nadu, India
Kavisankaar@gmail.com,drcc@annauniv.edu

**ABSTRACT**

*The challenging number is used for the detection of Spoofing attack. The IP Spoofing is considered to be one of the potentially brutal attack which acts as a tool for the DDoS attack which is considered to be a major threat among security problems in today's internet. These kinds of attack are extremely severe. They bring down business of company drastically. DDoS attack can easily exhaust the computing and communication resources of its victim within a short period of time. There are attacks exploiting some vulnerability or implementation bug in the software implementation of a service to bring that down and some attacks will use all the available resources at the target machine. This deals on attacks that consume all the bandwidth available to the victim machine. While concentrating on the bandwidth attack the TCP SYN flood is the more prominent attack. TCP/IP protocol suite is the most widely used protocol suite for data communication. The TCP SYN flood works by exhausting the TCP connection queue of the host and thus denying legitimate connection request. There are various methods used to detect and prevent this attack, one of which is to block the packet based on SYN flag count from the same IP address. This kind of prevention methods becomes unsuitable when the attackers use the Spoofed IP address. The SYN spoofing becomes a major tool the TCP SYN flooding. For the prevention of this kind of attacks, the TCP specific probing is used in the proposed scheme where the client is requested challenging number while sending the ACK in the three way hand shake. This is very useful to find the Spoofed IP Packets/TCP SYN flood and preventing them.*

## 1. INTRODUCTION

The SYN flood attack exploits a vulnerability of the TCP three-way handshake, namely, that a server needs to allocate a large data structure for any incoming SYN packet regardless of its authenticity. During SYN flood attacks, the attacker sends SYN packets with source IP addresses that do not exist or are not in use. This kind of flooding the SYN packet by spoofing their identity is SYN spoofing. During the three-way handshake, when the server puts the request information into the memory stack, it will wait for the confirmation from the client that sends the request. While the request is waiting to be confirmed, it will remain in the memory stack. Since the source IP addresses used in SYN flood attacks can be nonexistent, the server will not receive confirmation packets for requests created by the SYN flood attack. Each half-open connection will remain on the memory stack until it times out, it will retransmit the SYN-ACK 5 times, doubling the time-out value after each retransmission. The initial time-out value is 3 seconds, so retries are attempted at 3, 6, 12, 24, and 48 seconds. More and more requests will accumulate and fill up the memory stack. Therefore, no new request, including legitimate requests, can be processed and the services of the system are disabled. Generally, the space for the memory stack allocated by the operating system is small, and even a small scale SYN flood attack can be disruptive. On the other hand, SYN floods can be also launched from compromised machines using spoofed IP address / genuine source IP addresses given these compromised machines using are configured to ignore the SYN/ACK packets from the target.

## 2. IP Spoofing Introduction

The attackers to hide or mask their true use the IP spoofing as a tool. IP spoofing is one of the most common forms of on-line camouflage. In IP spoofing, an attacker gains unauthorized access to a computer or a network by making it appear that a malicious message has come from a trusted machine by "spoofing" the IP address of that machine.

### 2.1. The IP Spoofing Techniques

The SYN flood attack exploits a vulnerability of the TCP three-way handshake, namely, that a server needs to allocate a large data structure for any incoming SYN packet regardless of its authenticity. During SYN flood attacks, the attacker sends SYN packets with source IP addresses that do not exist or are not in use. During the three-way handshake, when the server puts the request information into the memory stack, it will wait for the confirmation from the client that sends the request. While the request is waiting to be confirmed, it will remain in the memory stack. Since the source IP addresses used in SYN flood attacks can be nonexistent, the server will not receive confirmation packets for requests created by the SYN flood attack. Each half-open connection will remain on the memory stack until it times out, it will retransmit the SYN-ACK 5 times, doubling the time-out value after each retransmission. The initial time-out value is 3 seconds, so retries are attempted at 3, 6, 12, 24, and 48 seconds. More and more requests will accumulate and fill up the memory stack. Therefore, no new request, including legitimate requests, can be processed and the services of the system are disabled. Generally, the space for the memory stack allocated by the operating system is small, and even a small scale SYN flood attack can be disruptive. On the other hand, SYN floods can be also launched from compromised machines using spoofed IP address / genuine source IP addresses given these compromised machines using are configured to ignore the SYN/ACK packets from the target.

### 2.2. IP Spoofing Today

From the results of the Spoofed Project [MIT Advanced Network Architecture Group 2007; Beverly and Bauer 2005] it is concluded that IP Spoofing continues to be a mammoth problem in today internet world. It is still a prospective tool used by malicious users for attack and misdirection. With existence of ingress/egress filters it may be concluded that attackers not able to spoof many addresses.

### 2.3. Botnets

A botnet is a collection of software agents, or robots, it is the malicious software Controlled using IRC bots. With the increase in the popularity of botnets, it is believe attackers no longer need to use IP spoofing. In fact, when considering botnets, IP spoofing remains a problem for defenders and an asset for attackers. In some botnet-based attacks, such as a DNS DDoS amplification attack, IP spoofing is vital to the attack's success. Even if botnets did not use IP spoofing, the threat of IP spoofing would still exist.

## 3. Related Works

The methods used for preventing TCP SYN flood is done by the following way using the server as the detector of the attack and the local router of the attacker is used to prevent the attack. To establish the TCP connection with the server, every client should send the SYN signal and have to respond the SYN/ACK signal with the ACK signal. To identify the attack, the SYN request sent by the client is stored in the server data table (database) until the acknowledgement from the client received by the server for the SYN ACK signal. The client information stored in the table is the IP address and the SYN count. If the count (number of SYN request sent by the attacker without the ACK signal to establish the connection) in the table exceeds the threshold limit, the victim intimate the details of the attack to its local router and it will be sent to the local router of the attacker to drop all the packets from the respective node. If suppose the local router compromise with the attacking node, that router is also prevented from throwing the packet.

OS Fingerprinting [13] can detect spoofing packets if the spoofed source can be actively fingerprinted. The resulting active fingerprint is different from the passively deployed fingerprint then it is considered to be the spoofing packet. Even then, results can be complicated by a firewall between the target and the spoofed source, if the firewall filters the fingerprinting probes, or alters the responses. Fingerprinting is not reliable enough to depend on.

Hop-Count Filtering (HCF) [10] observes the hop-count of packets arriving at a given host/server. During normal times measurement is made where, HCF creates a mapping of IP addresses to hop counts. Then, if an attacker sends a spoofing packet to the host; it is likely the hop-count of the packet will not match the expected hop-count for packets from the spoofed source address. Because legitimate hop-counts may change due to routing changes, strictly filtering all packets that do not match would lead to false positives. In order to minimize false positives, HCF only begins filtering traffic if some threshold amount of packets does not match their expected hop counts.

The method used to prevent the opening of connections to spoofed source addresses is SYN cookies [11]. When a server uses SYN cookies it does not allocate resources to a connection until the 3-way TCP handshake completes. First the server sends a SYN + ACK packet with a specially encoded initial sequence number, or cookie, that includes a hash of the TCP headers from the client's initial SYN packet, a timestamp, and the client's Maximum Segment Size (MSS). Then when it receives the client's response, the server can check the sequence number and create the necessary state only if the client's sequence number is the cookie value plus one. Because the cookie uses a hash involving the server's secret key, attackers should not
be able to guess the correct cookie values. However, because of performance concerns and some incompatibilities with TCP extensions, such as large windows, operating systems generally do not activate the SYN cookie mechanism until the host's SYN queue fills up. An attacker sending spoofing traffic at a low rate may avoid triggering the SYN cookie mechanism. Administrators may be able to forcibly enable SYN cookies for all connections, but should be aware of the side effects.

Another mechanism used for IP Spoofing prevention is using IP puzzles [12]. It provides active defense against spoofing. Here the server sends an IP puzzle to a client, and then the client needs to "solve" the puzzle by performing some computational task. Only after the server receives the puzzle solution from the client will the server allow the client to connect. It is prohibitively expensive for malicious hosts to send large numbers of packets as a side effect it is preventing attackers from successfully sending spoofing packets. Since the IP puzzle would be sent to the listed source and not the attacker, an attacker could not send a puzzle solution, thus preventing the attacker from spoofing.

## 4. CNoA using TCP Probing

CNoA method is used for the mitigation of TCP SYN flood with IP spoofing. Here the simple TCP handshake is vulnerable to prevent attackers from spoofing TCP packet, since attackers may be able to predict TCP sequence numbers. TCP-specific probes intelligently craft /append TCP acknowledgment messages to add another layer of protection. Since the sender of spoofing packets is often unable to see any replies, a recipient host can send acknowledgments that should reply with the challenging number, and then observe whether or not the supposed source responds correctly. If the supposed source does not respond with correct challenging number in acknowledgment packet, the recipient host considers the packet's source to be spoofed.

The recording of the TCP handshake is the key aspect in this method. The TCP acknowledgment message sends along with the SYN+ACK packet send from the victim server which undergoes the IP Spoofing for the TCP SYN flooding. If the packet is responded with correct challenging number the packet is not spoofed.

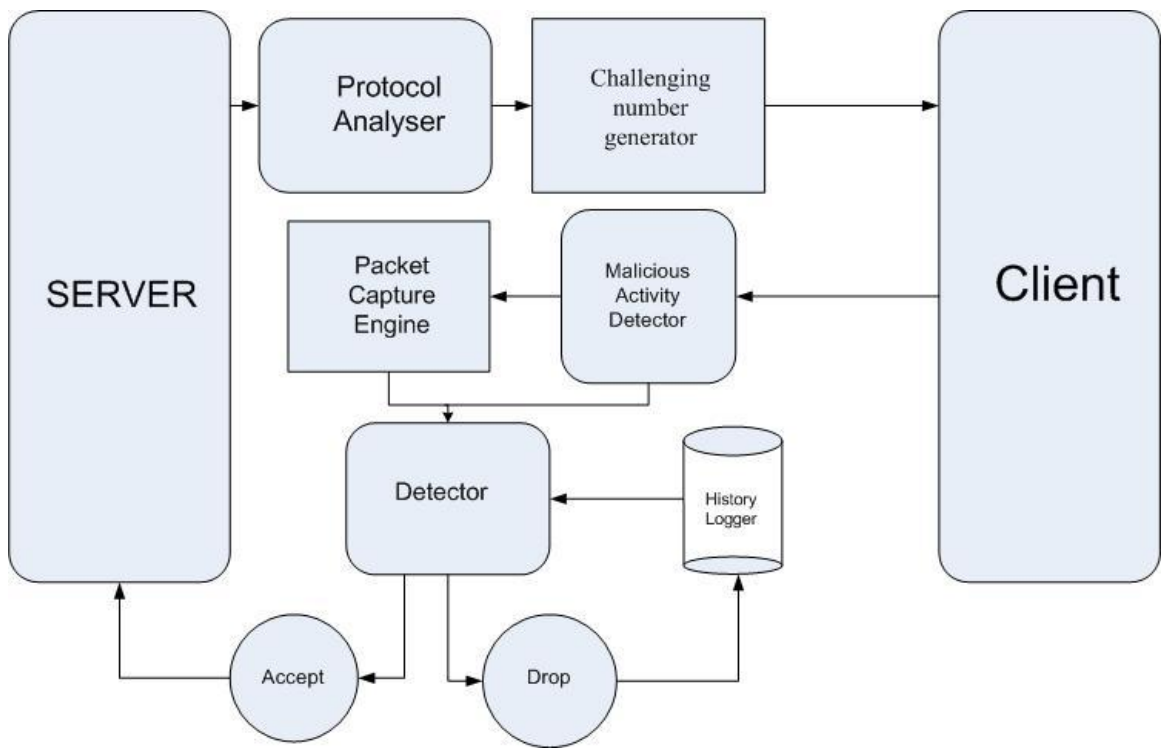

**Figure . 1.** .TCP SYN flood detection model using CNoA

From the (Figure 1) we can see the CNoA method gives a very good picture whether the TCP packet received is spoofed or not. Based on the detection the packet is dropped or accepted. Initially the server receives the TCP packet with SYN flag in the packet so the protocol analyzer detect it as TCP protocol and send the client with the TCP Acknowledgement if the packet does not come from the spoofed IP address then the client reply with correct Challenging number based on this reply the learning/recording packet record this and send it to the detector. Based on the reply from the client the packet is dropped or accepted.

This is a host-based architecture and it is developed using TCP Probing .Here the TCP probe is used to send the specification to the client trying to establish a new connection with the server. The decision is taken based on the reply from the client. CNoA is used to send the Challenging number when the client trying to establish a new connection with the server. The Specification contains the TCP probe which contains Challenging number that should make the client to send the same challenging number. The Protocol analyzer analyses the packet whether it is the TCP packet. The Packet Capture engine is used to record the packet used in the TCP handshake and it is useful in verifying the specification given by the TCP Probe. The Decider based on the information available (i.e.) it follows the Specification while replying back to the server, based on this packet accepted or rejected . The information of dropped packet is logged in the History logger.

## 5. Algorithm

```
if SYN packet from client IP then
{
    send SYN+ ACK+ Challenging Number from server

            if ACK + Challenging Number then
            {
                    TCP Connection Established
            }
            else
    {
                    Decision (History Logger)
    }
}
```

**Figure. 2.** TCP SYN Spoofing detection algorithm using CNoA

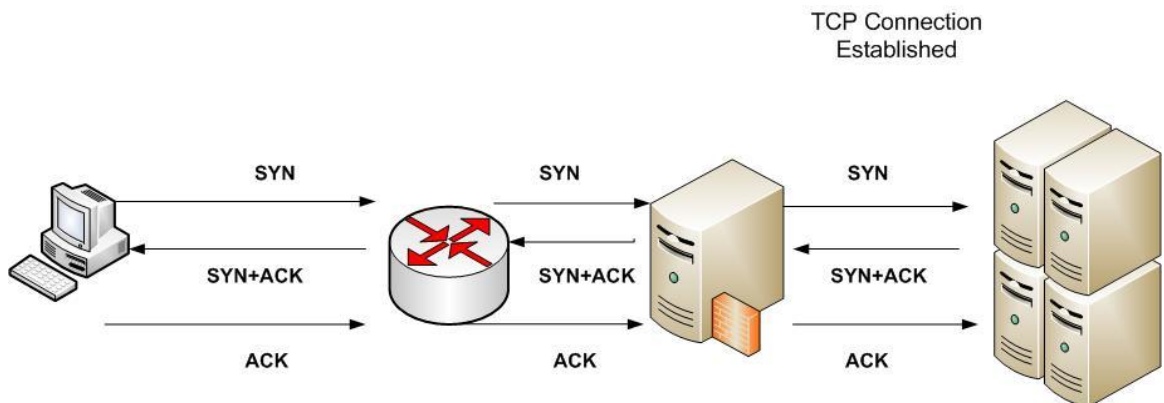

**Figure. 3.** TCP Three way handshake in normal scenario

The above diagram represents the TCP Connection establishment where the TCP Three way handshake takes place. Here this protocol have the risk of TCP SYN flooding where the SYN packets are flooded continuously makes it's a Denial of service attack by consuming the bandwidth of the network. It is made worse by SYN spoofing i.e. changing the identity of the attacker. It becomes impossible to prevent and attack from the spoofed source.

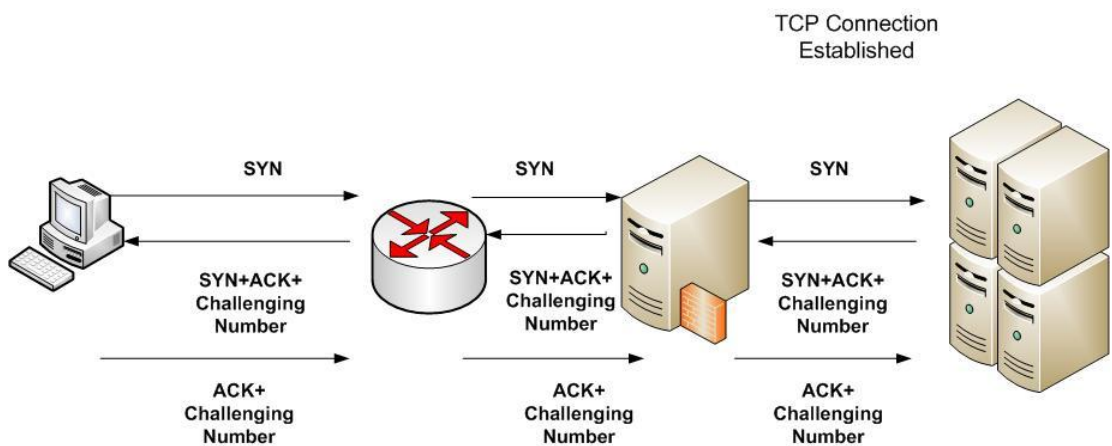

**Figure. 4.** TCP Three way handshake connection established using CNoA method

The CNoA method is used in the TCP Three way handshake. The server generators the Challenging number which is send along with the SYN+ACK i.e. (SYN +ACK+ Challenging number) for which the clients as to reply the ACK with the challenging number i.e. (ACK+ Challenging number) send by the server. Here the server verifies the challenging number if it matches the connection is established. If the challenging number does not match or it doesn't send the replay packet connection is not established.

## 5. Test Cases

The attack module is done from the Blade Server and the Storage Server run the prevention of the flooding attack. We are using two different systems for testing the client and the server program respectively.

**Table. 1.** Attack, detection and prevention model of CNoA method

|  | IPv4 | IPv6 |
|---|---|---|
| Attack | TCP SYN flooding with SYN Spoofing | TCP SYN flooding with SYN Spoofing |
| Detection | A challenging number for connection establishment | A challenging number for connection establishment |
| Prevention | Add the IP address in black list | Add the IP address in black list |

## Case 1: TCP Connection Establishment Using IPv6

The client establishes connection with the server using 3-way handshake with IPv6 global address. The server generates the challenging number to send to the client while establishing the connection with the server in the TCP three way handshakes.

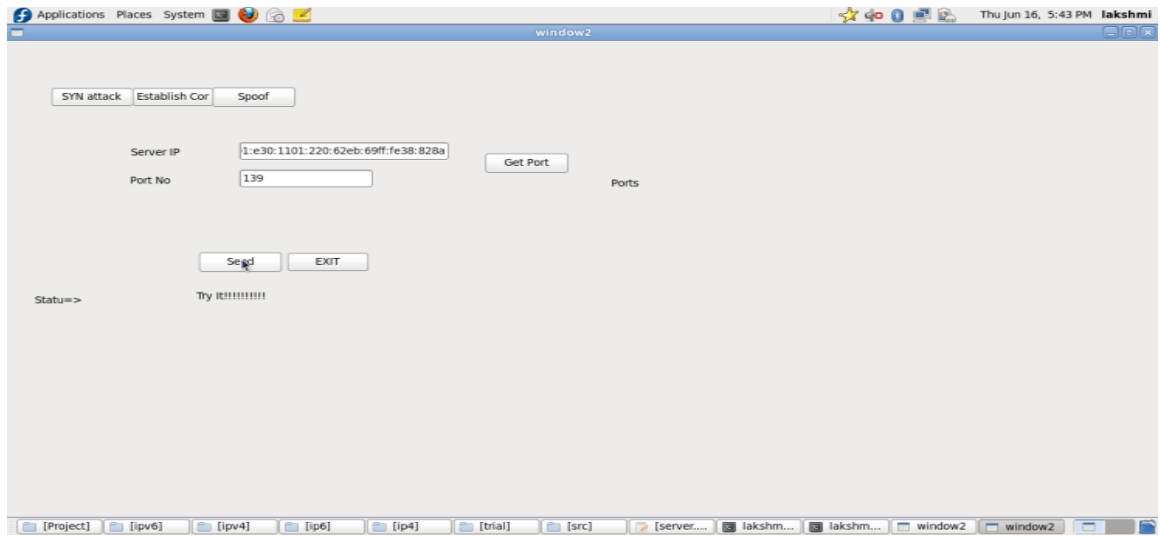

**Figure. 5. 1.** Connection Establishment in the Client

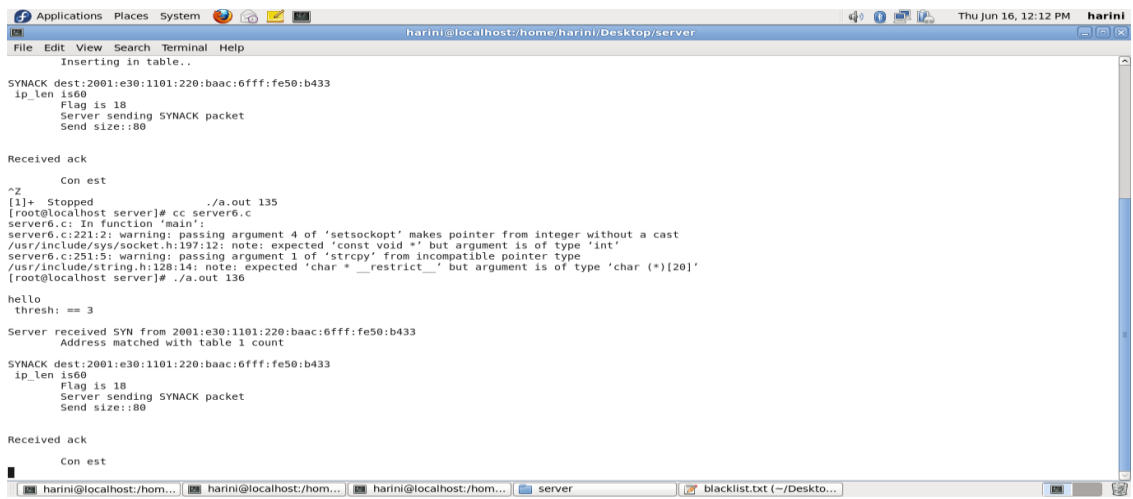

**Figure. 5. 2.** Connection Establishment in the Server

Using the challenging number generated by the server the client establishes connection with the server.

## Case 2: TCP IP Spoofing Detection and Prevention Using IPv6

The attacker tries to establish connection with the server. The server checks the ACK packet sent by the client to find out if it has agreed upon the challenging number that it has send to the client. It establishes the connection only if it had satisfies the above condition. Otherwise, it terminates the connection.

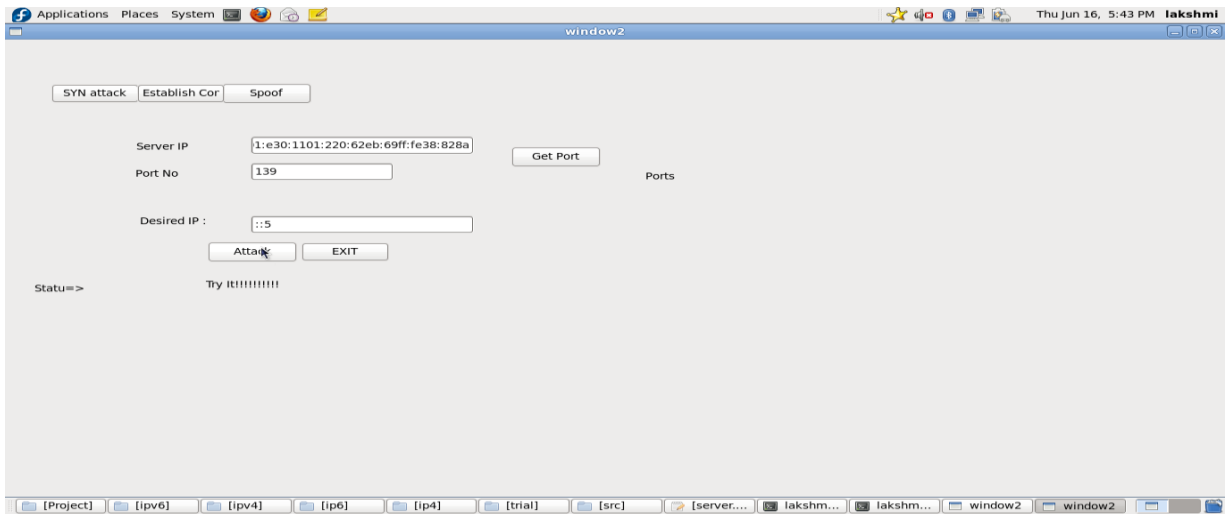

**Figure. 6. 1.** SYN Spoofing attack used for TCP SYN Flooding

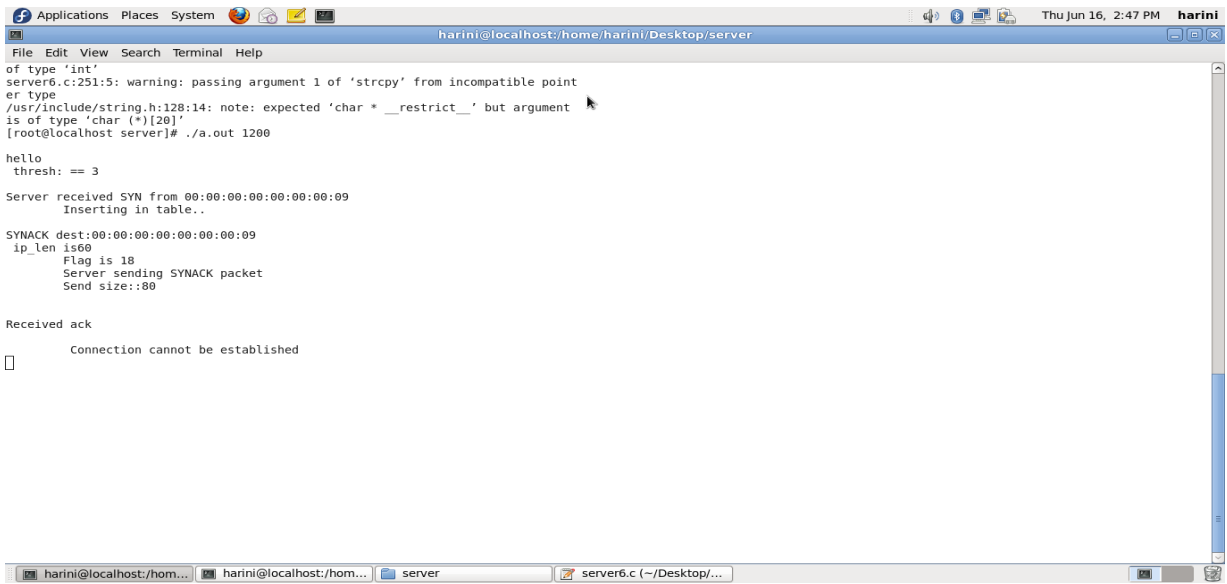

**Figure. 6. 2**. SYN Spoofing Detection and Prevention method using the CNoA

## Case 3: TCP SYN Flooding Attack Using IPv4

The server is flooded with SYN packet from the attacker using the IPv4 address.

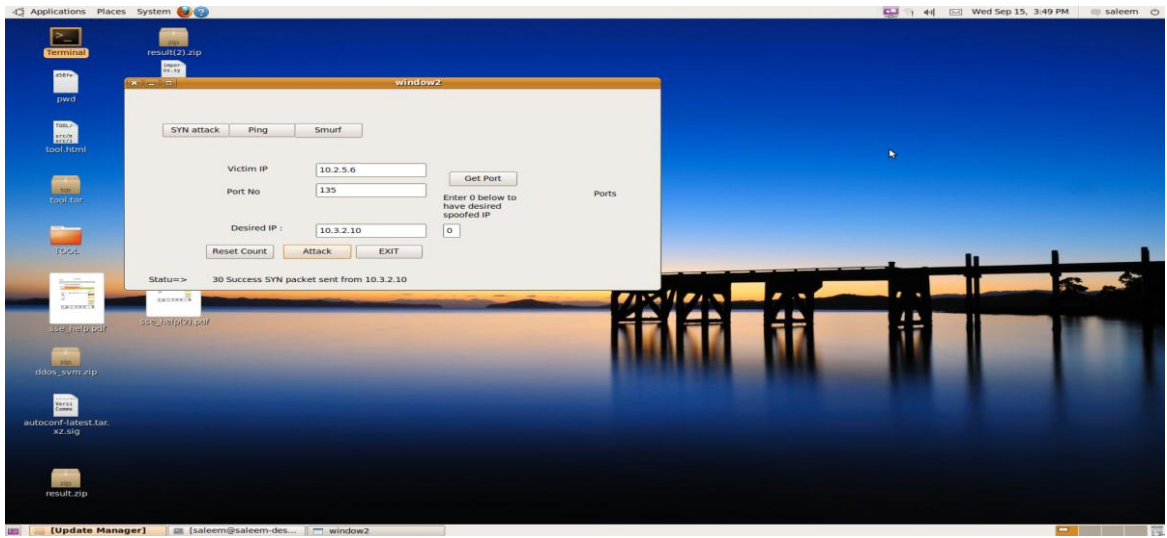

**Figure. 7. 1.** TCP SYN Flooding Attack using IPv4 address

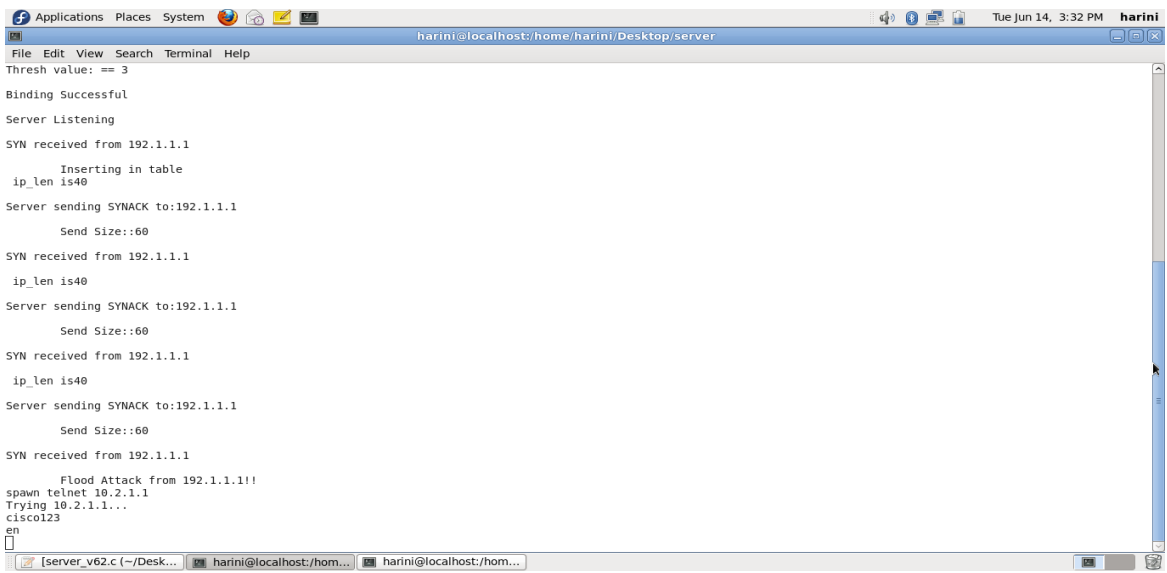

**Figure. 7. 2.** TCP SYN Flood Attack Prevention for IPv4 address

The CNoA method finds the spoofed Packet and drop the packet based on the speciation given in the acknowledgement. Here the comparison is made between the normal flow of packet with spoofed IP and the No. of spoofed packet detected using the CNoA. The attack is done on the blades servers and storage server maintains the challenging number method finds and detects the spoofed IP address and also drops the spoofed packets.

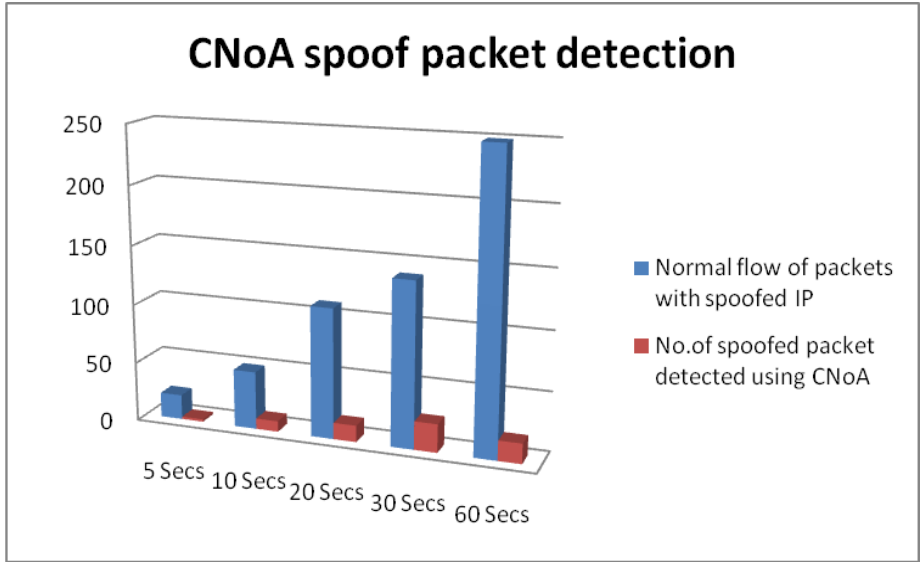

**Figure. 8.** Comparison of spoofed packet in normal flow and the spoofed packet detected using CNoA method

The CNoA method finds the spoofed Packet and drop the packet based on the speciation given in the acknowledgement. Here the comparison is made between the normal flow of packet with spoofed IP and the No. of spoofed packet detected using the CNoA method.

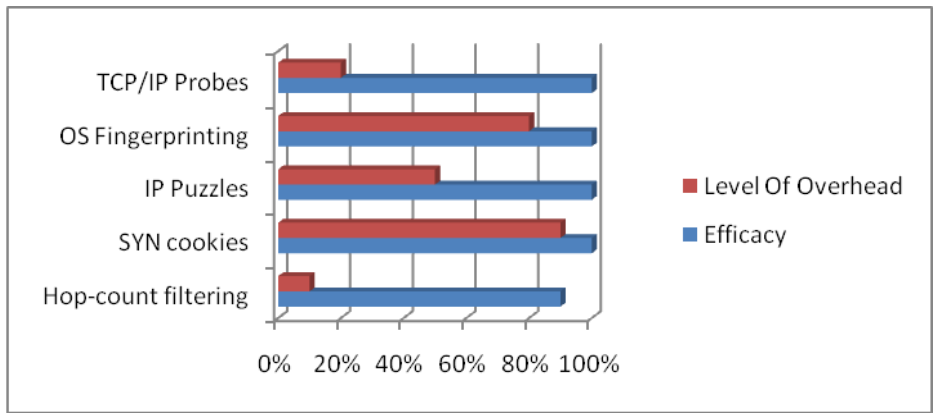

**Figure. 9.** Comparison of various IP Spoofing methods with CNoA method

Figure. 9. The TCP/IP Probe seems to have better efficiency with less overhead as compared to other methods of IP Spoofing prevention for the purpose of DDoS attack protection.

## 6. CONCLUSIONS

The CNoA method is very much useful in detecting SYN Spoofing and prevention of the TCP SYN flooding attack. The TCP SYN flooding which proves to be unpreventable by spoofing their identity is mitigated by using this method. Here the computational overhead of appending the challenging number during the connection establishment is minimized.

The solution is given in the server-level i.e. host based solution is only given. The spoofed packets are to found in the router itself so that the bandwidth consumed to transfer the malicious data from source to destination can be avoided.

## Acknowledgements

This Work is supported by the NTRO, Government of India. NTRO provides the fund for collaborative project "Collaborative Directed Basic Research on Smart and Secure Environment" and this paper is modelled for this project. Authors would like to thank the project coordinators and the NTRO officials.

**Authors**

L. Kavisankar is a PhD student in the Department of Computer Science and Engineering at Anna University, Chennai, India. He received his B.E Computer Science and Engineering from Easwari Engineering College under Anna University in 2007 and M.E Computer Science and Engineering from SSN College of Engineering Under Anna University in 2009, Chennai, India. His current research is on mobile IPv6 and network security

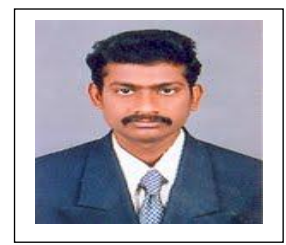

Dr. C. Chellappan is a Professor in the Department of Computer Science and Engineering at Anna University, Chennai, India. He received his B.Sc. in Applied Sciences and M.Sc in Applied Science–Applied Mathematics from PSG College Technology, Coimbatore under University of Madras in 1972 and 1977. He received his M.E and Ph.D in Computer Science and Engineering from Anna University in 1982 and 1987. He was the Director Of Ramanujan Computing Centre (RCC) for 3 years at Anna University (2002–2005). He has published more than 70 papers in reputed International Journals and Conferences. His research areas are computer networks, Distributed / mobile computing and soft computing, software agent, object oriented design and network security.

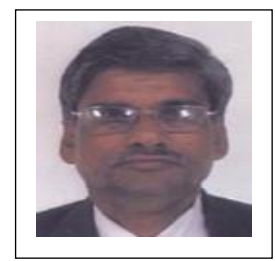